\documentclass[a4paper,11pt]{article}
\pdfoutput=1 

\usepackage{jcappub} 

\usepackage[T1]{fontenc} 
\usepackage{graphicx}	
\usepackage{amsmath}	
\usepackage{bm}
\usepackage{mathrsfs}
\usepackage{latexsym}
\usepackage{float}
\usepackage{color}
\usepackage[normalem]{ulem} 
\usepackage{dcolumn}
\usepackage[usenames,dvipsnames]{xcolor}
\usepackage[caption=false]{subfig}
\usepackage{slashed}
\usepackage{multirow}

\DeclareUnicodeCharacter{2212}{-}

\title{Radial oscillations in neutron stars from unified hadronic and quarkyonic equation of states}

\author[a]{Souhardya Sen}
\author[b]{,Shubham Kumar}
\author[a]{,Athul Kunjipurayil}
\author[a]{,Pinku Routaray}
\author[c]{,Tianqi Zhao}
\author[a,1]{and Bharat Kumar \note{Corresponding author.}}

\affiliation[a]{Department of Physics and Astronomy, National Institute of Technology,\\Rourkela 769008, India}
\affiliation[b]{Department of Physical Sciences, Indian Institute of Science Education and Research Kolkata,\\Mohanpur 741 246, WB, India}
\affiliation[c]{Department of Physics and Astronomy, Ohio University, \\Athens, OH~45701, USA}

\emailAdd{souhardya147@gmail.com}
\emailAdd{acad.shubham@gmail.com}
\emailAdd{mywhatsappkp@gmail.com}
\emailAdd{routaraypinku@gmail.com}
\emailAdd{zhaot@ohio.edu}
\emailAdd{kumarbh@nitrkl.ac.in}

\abstract{We study radial oscillations in non-rotating neutron stars by considering the unified equation of states (EoSs), which support the 2 M$_\odot$ star criterion. We solve the Sturm-Liouville problem to compute 20 lowest radial oscillation modes and their eigenfunctions for neutron star modelled with eight selected unified EoSs from distinct Skyrme-Hartree Fock, Relativistic Mean-Field and quarkyonic models. We compare the behavior of the computed eigenfrequency for NS modelled with hadronic to that with quarkyonic EoSs while varying central densities. The lowest order, f-mode frequency varies substantially between the two classes of the of EoS at 1.4 M$_\odot$ but vanishes at their respective maximum masses, consistent with the stability criterion $\partial M/\partial\rho_c > 0$. Moreover, we also computed large frequency separation and discovered that higher-order mode frequencies are significantly reduced by incorporating crust in the EoS.}


\arxivnumber{2205.02076}

\begin{document}
\maketitle
\flushbottom

\section{Introduction}
\label{sec:intro}
Neutron stars (NS) are the collapsing cores of previously massive stars that form after supernova explosions \cite{PhysRev.46.76.2}. A NS's central density is predicted to be around ten times that of the nuclear saturation density ($\rho_0\approx$ 0.16 fm$^{-3}$). Some unusual phases, such as meson condensation \cite{migdal1972stability, Migdal_1972, mannarelli2019meson} or quark deconfinement \cite{rajagopal2001condensed, alford2008color, anglani2014crystalline}, can be achieved at such high densities. It is impossible to achieve these circumstances on Earth. As a result, studying NS gives us unique insights into the physics of strongly interacting nuclear matter and phase transitions at ultra-high densities. Understanding NS also necessitates knowledge of several scientific fields like nuclear physics, particle physics, astrophysics as well as gravitational physics. Even while we know a lot about how an NS forms, we know relatively little about its internal composition. The one known fact in NS is that characteristics like its mass and radius are strongly influenced by the equation of state (EoS) of dense matter. Our overall objective in this field is to identify such properties that may be observed so that we can utilise the generated observational data to construct an accurate EoS, thereby bringing us one step closer to completely comprehending the interior of a NS.

NSs are shown to pulse with different quasi-normal modes (QNMs) in which infinitesimal perturbations induce oscillations whose amplitude decays exponentially with time due to various damping mechanics. Dynamical instabilities like mass accretion and tidal forces from a nearby binary companion \cite{Chirenti_2017, PhysRevLett.116.181101}, star-quakes generated by fissures in the crust \cite{Franco_2000}, and supernova explosions are a few sources that might induce oscillations inside an NS \cite{PhysRevLett.108.011102}. QNMs are categorised into two main groups based on the motion of the pulsations: radial and non-radial oscillation, and each of them is further classified based on the restoring force that acts on the displaced mass element to bring the system back to equilibrium. These restoring forces maybe gravity ($g$-mode), pressure gradient ($p$-mode), Coriolis forces ($r$-mode), magnetic fields, and centrifugal and Coriolis forces in rotating NS, distinguished by their frequency range. Here, we focus on pressure being our restoring force.

Because QNMs are sensitive to the internal composition and EoS of dense matter, we analyse the internal structure of the star and identify the thermodynamic parameters of the NS's interior using asteroseismology for the study of stellar pulsations in general relativistic frame. The study of frequencies can thus assist us in indirectly probing within an NS and discovering how strongly interacting nuclear matter behaves at such high densities, thereby further constraining the choice of EoS \cite{brillante2014radial, kokkotas2001radial, miniutti2003non, PhysRevD.96.083013, passamonti2005coupling, passamonti2006coupling, refId0S, PhysRevC.96.065803, flores2010radial, 1992A&A...260..250V}.

Although radial oscillations cannot directly emit gravitational waves (GWs), they can couple with non-radial oscillations, amplifying them and creating a stronger GW that might be detected \cite{passamonti2006coupling, PhysRevD.75.084038}. During the formation of hyper-massive NS through binary NS merger, a short gamma-ray burst (SGRB) is emitted, which can be modulated by the radial oscillations \cite{Chirenti_2019}. As a result, it is not only valuable in understanding the physics of dense nuclear matter inside an NS, but it also has some application in GW physics.

Chandrasekhar investigated radial oscillations in NS, being the simplest mode of oscillation, for the first time in 1964 \cite{1964ApJ...140..417C}. Following that, it was researched by other authors such as Harrison et al. \cite{1965gtgc.book.....H} and Chanmugam \cite{chanmugam1977radial}, employing zero temperature EoS and finite temperature EoS for Proto-NS by Gondek et al. \cite{gondek1997radial}, and strange stars by Benvenuto \& Horvath \cite{benvenuto2013evolution}, Gondek \& Zdunik \cite{gondek1999avoided}. Glass and Lindblom performed the first major investigation of radial oscillations in 1983 \cite{1983ApJS...53...93G}. Their numerical results were later rectified by Väth \& Chanmugam in 1992 \cite{1992A&A...260..250V} and re-examined by Kokkotas \& Ruoff \cite{kokkotas2001radial} using two alternative numerical approaches that included six more zero temperature EoS. Their results suggested that oscillations become unstable after the central density at which NS reaches its maximum mass. This was due to the fact that they used the equilibrium adiabatic index in all of their equations. However, if different adiabatic indices connected to the physical circumstances inside NS are employed \cite{gondek1997radial} and the slowness of weakly interacting processes is taken into account, stability can be extended beyond that central density \cite{chanmugam1977radial, gourgoulhon1995maximum}.

The radial modes of oscillation are calculated in this work using eight unified EoSs based on the Relativistic Mean Field (RMF), Skyrme-Hartree-Fock (SHF) \cite{Fortin,vishal} and quarkyonic models \cite{PhysRevD.102.023021}. We solve the Sturm-Liouville eigenvalue problem \cite{diffeq, diffeq2, 1980tsp..book.....C} with the assumption that our NS is non-rotating and has a zero magnetic field. We base our theory on the fact that the oscillations are small enough to use linear theory. Also, they are adiabatic such that damping time scale is much longer than oscillation period \cite{1966ApJ...145..505B, 1989A&A...217..137H}.

This work is structured as follows: in the upcoming section \ref{sec:form}, we go over the theoretical formalism, starting with hydrostatic equilibrium and stellar structure equations in general relativity in sub-section \ref{subsec:TOV}, followed by radial oscillation equations in sub-section \ref{subsec:radosc}. We discuss our numerical approach in sub-sub-section \ref{subsec:num}. In section \ref{sec:EoS}, we offer a quick summary of our chosen eight EoS and the reasons for selecting them. In section \ref{sec:R&D}, we provide our numerical results. Sub-section \ref{subsec:mr} corresponds to the computation of mass-radius, whereas in  sub-section \ref{subsec:oscmode}, we present our calculated radial mode frequencies in order to analyse them, followed by their chances and ways of detection in sub-section \ref{subsec:detect}. Finally, we summarise and conclude, as well as suggest opportunities for further improving our current work, in section \ref{sec:con}.

\section{Theoretical formalism}
\label{sec:form}
The massive gravitational forces of a non-rotating NS's interior allow only slight deviations from spherical symmetry, resulting in a nearly perfect formed sphere in its equilibrium condition. As a result, our assumption that the star is spherically symmetric is a reasonable approximation. The gravitational field of such a body is itself spherically symmetric and is given by the Schwarzschild metric in the form of \cite{1916AbhKP1916..189S}:
\begin{equation}
ds^2 = -e^{2\nu} c^2 dt^2 + e^{2\lambda} dr^2 + r^2(d\theta^2 + \text{sin}^2\theta d\phi^2),
\label{eqn4}
\end{equation}
where $\lambda \equiv \lambda (r)$ and $\nu \equiv \nu (r)$ following their own set of equations. Here, the energy-momentum tensor $T_{\mu\nu}$ has the form of a perfect fluid:
\begin{equation}
T_{\mu\nu}=(P+\mathcal{E})u_\mu u_\nu + Pg_{\mu\nu},
\label{eqn14}
\end{equation}
where $P$ is the pressure, $\mathcal{E}$ represents the energy density and $u_\mu$ is the covariant velocity. Since we have spherical symmetry and are only going to consider motion along radial direction, only the components $u_0$ and $u_1$ are non-zero.\footnote{Here, we use the mostly positive signature $(- + + +)$.}

\subsection{Hydrostatic equilibrium equations}
\label{subsec:TOV}
In a state of hydrostatic equilibrium, all quantities are time-independent. Therefore, even $u_1$ is zero. From Einstein's field equations, using the Schwarzschild metric in eq. (\ref{eqn4}) in equilibrium and applying the boundary condition $\lambda(r=0)=0$, we get:
\begin{equation}
e^{-2\lambda(r)} = \left(1-\frac{2Gm}{c^2r}\right),
\label{eqn7}
\end{equation}
where the mass $m$ can be integrated using:
\begin{equation}
\frac{dm}{dr} = \frac{4\pi r^2 \mathcal{E}}{c^2} .
\label{eqn6}
\end{equation}
Similarly, using the law of conservation of momentum, we get \cite{LandauLif}:
\begin{equation}
\frac{d\nu}{dr}=-\frac{1}{P+\mathcal{E}}\frac{dP}{dr}.
\label{eqn8}
\end{equation}
Finally, using eq. (\ref{eqn8}) and the Einstein's field equations, we get:
\begin{equation}
\frac{dP}{dr} = - \frac{Gm}{c^2r^2} \frac{\left( P+\mathcal{E} \right) \left( 1 + \frac{4\pi r^3P}{mc^2}\right)}{\left(1-\frac{2Gm}{c^2r}\right)}
\label{eqn5}
\end{equation}
 Eqs. (\ref{eqn7}) and (\ref{eqn8}) define the behaviour of the metric functions inside the NS where $r<R$. At the surface, i.e. at $r=R$, they satisfy the boundary condition, 
\begin{equation}
e^{2\nu(R)} = e^{-2\lambda(R)} = \left(1-\frac{2Gm}{c^2R}\right).
\label{eqn9}
\end{equation}
Eq. (\ref{eqn9}) stays true even outside the star, where $R$ should be replaced by $r$ for $r>R$ as it attains the familiar form of the Schwarzschild solution. 

Eqs. (\ref{eqn6}) and (\ref{eqn5}) are collectively known as the Tolman–Oppenheimer–Volkoff (TOV) equations \cite{oppenheimer1939massive, tolman1939static}. These equations express the equilibrium at each step of the radius $r$, between the internal pressure of the overlying material against the gravitational force of attraction. These equations can be interpreted if we consider a shell of radius $r$ and thickness $dr$, with a pressure difference of $dp$ in the exterior with respect to its interior and evaluate the net force on each side.
The only thing needed to solve the structure equations of NSs is the Equation of State (EoS) of dense matter, i.e., the relation between the pressure and the energy density, which enters the TOV equations. For a given EoS, the TOV equations can be integrated from the origin with the boundary conditions $m(r=0)=0$ and $P(r=0) = P_c$, where $P_c$ is the central pressure, until the pressure becomes zero. The point $R$, where the pressure vanishes, provides the circumferential radius of the star. The integration of eq. (\ref{eqn6}) from zero to $R$ gives its total mass $m(R)=M$.

\subsection{Radial oscillation equations}
\label{subsec:radosc}

Keeping the spherical symmetry of the background equilibrium configuration, we perturb both fluid and spacetime variables. Assuming a harmonic time dependence for the radial displacement of the fluid element located at position $r$ in the unperturbed model
\begin{equation}
    \delta r (r,t) = X(r)e^{i\omega t},
\end{equation}
the linearized radial perturbation equations can be written as \cite{kokkotas2001radial}
\begin{equation}\label{RPE}
\begin{aligned}
c_{s}^{2} X^{\prime \prime}+\left(\left(c_{s}^{2}\right)^{\prime}-Z+ \frac{4 \pi G}{c^4} r \gamma P e^{2 \lambda}-\nu^{\prime} c^2 \right) X^{\prime}& \\
+ {\left[2\left(\nu^{\prime}\right)^{2}c^2+\frac{2 G m}{r^{3}} e^{2 \lambda}-Z^{\prime}-\frac{4 \pi G }{c^4} (P+\mathcal{E}) Z r e^{2 \lambda} + \omega^{2} e^{2 \lambda-2 \nu}\right]X} &= 0,
\end{aligned}
\end{equation}
where primes denote differentiation with respect to radial coordinate $r$ and $c_{s}^{2}=\frac{dP}{d\mathcal{E}}$ is the speed of sound squared in units of $c^2$. $\gamma$ is the adiabatic index, related to the speed of sound by
\begin{equation}
    \gamma=\left(\frac{P+\mathcal{E}}{P}\right)c_{s}^{2},
\end{equation}
and
\begin{equation}
    Z(r)=c_{s}^{2}\left(\nu^{\prime}-\frac{2}{r}\right).
\end{equation}
Now, we re-define the displacement function as
\begin{equation}\label{zeta}
    \zeta=r^{2}e^{-\nu}X.
\end{equation}
Eq. (\ref{RPE}) can be rewritten for $\zeta$ as
\begin{equation}\label{RPEs}
\frac{d}{dr}\left( H\frac{d \zeta}{d r}\right)+\left(\omega^{2} W+Q\right) \zeta=0,
\end{equation} 
with
\begin{subequations}\label{rf}
\begin{align}
\label{rf1}
H&=r^{-2}(P+\mathcal{E}) e^{\lambda+3 \nu} c_{s}^{2} \\
\label{rf2}
W&=r^{-2}(P+\mathcal{E}) e^{3 \lambda+\nu} \\
\label{rf3}
Q&=r^{-2}(P+\mathcal{E}) e^{\lambda+3 \nu}\left(\left( \nu^\prime \right)^{2}+\frac{4}{r} \nu^{\prime}- \frac{8 \pi G}{c^4} e^{2 \lambda} P\right) .
\end{align}
\end{subequations}

$H$, $W$, and $Q$ are functions of radial coordinate $r$ and can be calculated using the unperturbed background configuration. Note, eq. \eqref{RPEs} explicitly shows its self-adjoint nature. The Lagrangian variation of
the pressure now takes the simple form
\begin{equation}\label{dp}
    \Delta P = -r^{-2} e^{\nu}(P+\mathcal{E}) c_{s}^{2}\zeta^{\prime}.
\end{equation}
The boundary condition at the center is 
\begin{equation}\label{BC1}
    X(r=0)=0,
\end{equation}
because the fluid element there can not be displaced for radial oscillations, and at the stellar surface, the Lagrangian variation of pressure should vanish
\begin{equation}\label{BC2}
    \Delta P(R) = 0.
\end{equation}

The differential equation \eqref{RPEs} subject to boundary conditions eqs. \eqref{BC1} and \eqref{BC2} is a Sturm-Liouville eigenvalue problem. The eigenvalues $\omega_{n}^{2}$ are real and form an infinite, discrete sequence with
\begin{equation*}
    \omega_{0}^{2} < \omega_{1}^{2} < \omega_{2}^{2} <...
\end{equation*}
The eigenfunction of the $n$-th mode has precisely $n$ number of zeros between the centre and the surface of the star. $\omega$ is real for $\omega^{2} > 0$, and thus the solution is purely oscillatory. However, for $\omega^{2} < 0$, we have an imaginary frequency, which corresponds to an exponentially growing solution. Since the general solution is always a superposition of all such solutions $\omega_{n}s$, the presence of an exponentially growing solution corresponds to instability in the radial oscillations. For NSs, the fundamental mode $\omega_{0}$ becomes unstable exactly at the central density $\rho_{c}$ greater than the $\rho_{\text{critical}}$ corresponding to the maximum mass configuration. The star will eventually collapse to a black hole in that case.

\subsubsection{Numerical method}
\label{subsec:num}
For numerical integration, we write eq. \eqref{RPEs} as a system of two first order differential equations in $\zeta$ and $\eta = H \zeta^{\prime}$:
\begin{equation}\label{DE1}
    \frac{d \zeta}{d r} = \frac{\eta}{H} 
\end{equation}
and
\begin{equation}\label{DE2}
    \frac{d \eta}{d r} = -\left(\omega^{2} W+Q\right) \zeta.
\end{equation}
Expanding $\zeta$ and $\eta$ close to the origin and comparing the leading order coefficients gives $\eta_0 = 3\zeta_0H_0$ \cite{kokkotas2001radial}. Here $\eta_0$, $\zeta_0$ and $H_0$ are their corresponding values at $r=r_{\text{min}}$, where $r_{\text{min}}$ is the smallest radial coordinate considered in the integration of the TOV equations. Choosing $\eta_0=1$, we get $\zeta_0=1 / (3H_0)$. At the surface, boundary condition eq. \eqref{BC2} implies
\begin{equation}\label{BCeta}
   \eta(R)=0. 
\end{equation}
For an arbitrary value of $\omega$, the integration can be done from the stellar centre to its surface, where $\eta(R)$ is obtained. Those $\omega$s for which eq. \eqref{BCeta} is satisfied, are the eigenfrequencies of the radial oscillation. 

\section{Equation of State}
\label{sec:EoS}
To solve the radial oscillation of the NS, we choose five unified hadronic EoSs based on RMF and SHF models with different parametrization \cite{Fortin,vishal} and three quarkyonic EoSs \cite{PhysRevLett.122.122701, PhysRevD.102.023021}, viz:
\begin{enumerate}
   \item {NL3:} The famous NL3 is based on non-linear interaction, where only the $\sigma$-meson self-coupling term is included while the cross-coupling terms are not taken into account \cite{PhysRevC.55.540}.
   \item{IOPB:} Interaction with higher-order couplings including self-coupling of $\rho$-mesons and $\omega$-$\rho$ cross-coupling terms \cite{kumar2018new}.
   \item{DD2:} A density-dependent interaction with experimental values of proton and neutron mass $m_p$,$m_n$. This model can provide an accurate description of the composition, and thermodynamic quantities over a large range of densities \cite{typel2010composition}.
   \item{DDME2:} Also, a effective mean-field interaction with density-dependent meson-nucleon couplings \cite{PhysRevC.DDME2}.
   \item{SLy4:} Based on the Skyrme-Lyon model, this interaction is suitable for calculating the properties of neutron-rich matter. This model can describe both the NS crust as well as the liquid core \cite{refId0}.
   \item{Q1-3:} A quark-to-hadron crossover transition model with leptons and nucleons coexist in quarkyonic phase and degenerate in momentum space. The chosen sets of parameters are as follows \cite{PhysRevD.102.023021}:
   \begin{itemize}
    \item {Q1:} $L$ = 30 MeV, $\Lambda$= 1400 MeV, $n_t$ = 0.3 fm$^{-3}$,
    \item {Q2:} $L$ = 30 MeV, $\Lambda$= 800 MeV, $n_t$ = 0.3 fm$^{-3}$,
    \item {Q3:} $L$ = 50 MeV, $\Lambda$= 1400 MeV, $n_t$ = 0.4 fm$^{-3}$,
    \end{itemize}
    where $L$ is the slope parameter, $\Lambda$ is the dimensionless tidal deformability and $n_t$ is the transition density between the nucleonic and quarkyonic phase.
\end{enumerate}

\begin{figure}
    \centering
    \includegraphics[scale=0.55]{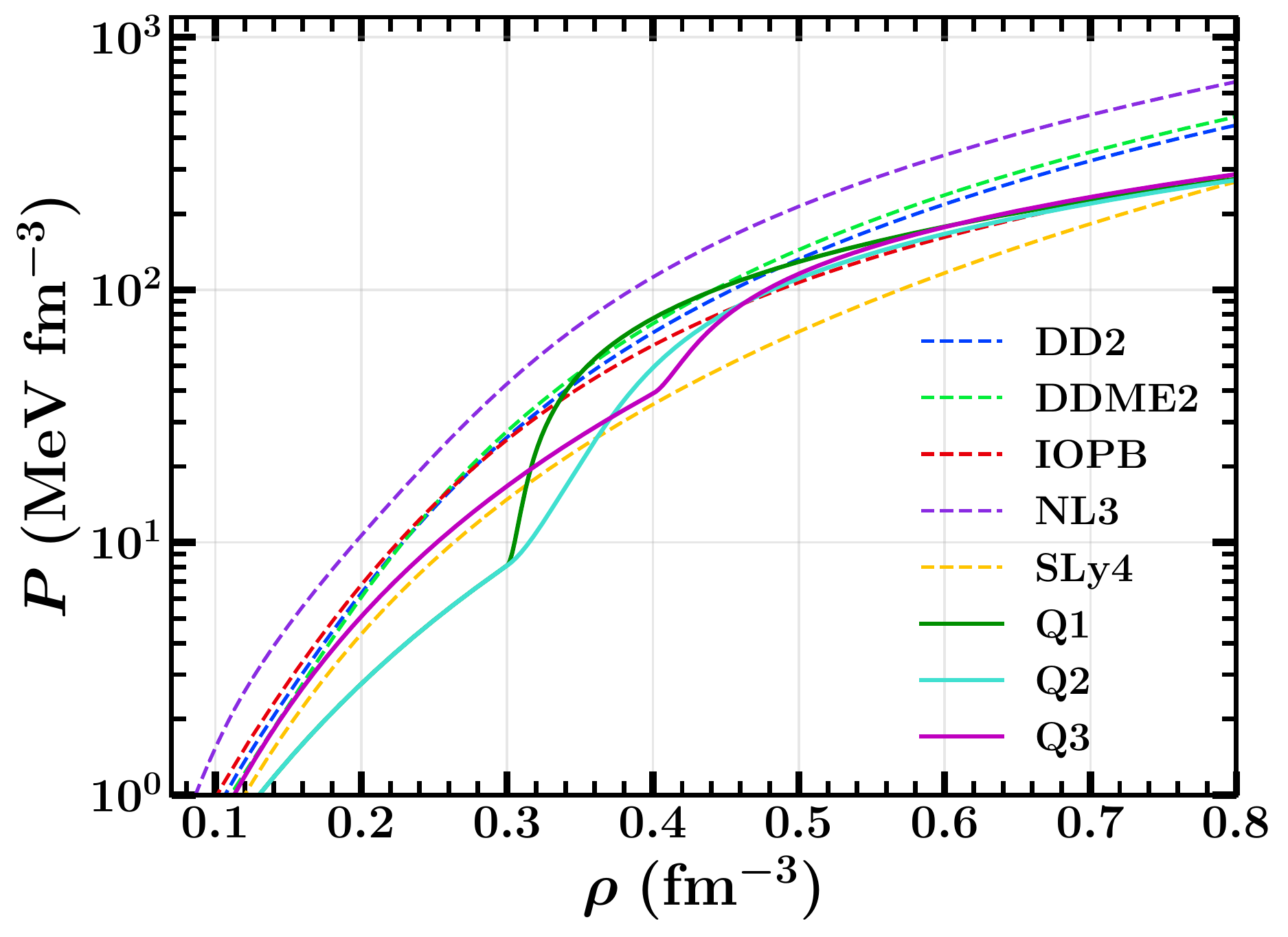}
    \caption{Pressure in NS matter vs baryon density calculated using five unified hadronic as well as three quarkyonic EoSs \cite{Fortin,vishal}}
    \label{fig:EoS}
\end{figure}

Figure \ref{fig:EoS} shows the pressure calculated with five unified hadronic EoSs NL3 \cite{PhysRevC.55.540}, IOPB \cite{kumar2018new}, DD2 \cite{typel2010composition}, DDME2 \cite{PhysRevC.DDME2}, SLy4 \cite{refId0} using dashed lines and with three quarkyonic EoSs Q1-3 \cite{PhysRevD.102.023021} using solid lines.

Among these chosen EoSs, NL3 yields the stiffer EoS, followed by DDME2 and DD2. SLy4 is initially the softest at lower densities but becomes stiffer than the remaining two EoSs as density rises. On the contrary, IOPB starts off stiffer than SLy4 but becomes softer with increasing density. At lower densities, Q1 and Q2 begin as softer EoSs, whereas Q3 is stiffer. These quarkyonic EoSs exhibit a phase transition at their respective transition densities, but they abruptly stiffen up when quarks flow off of nucleons and fill the lower momenta states, causing a spike in the pressure. We will use above mentioned EoSs for the calculation of radial oscillations and NS properties.

\section{Results and Discussion} 
\label{sec:R&D}
In the relativistic frame, the radial modes and global properties for eight separate sets, each with its own EoS, are computed concurrently. Runge-Kutta method of fourth order is employed to solve the set of coupled first order differential equations (\ref{eqn6}-\ref{eqn5}) employing a logarithmic step size ranging from 6 m near the centre to about 1 m near the surface. This unequal variation aids in recording the rapid change in $\gamma$ near the surface. The solution is iterated until the boundary conditions are fulfilled. Using these solutions, we solve another set of coupled first order differential equations (\ref{DE1}-\ref{DE2}) using eq. \eqref{rf}. Here, the boundary condition, eq. \eqref{BCeta}, is associated with a change of sign, thus can be calculated numerically by root finding methods such as the bisection method or the shooting method. We have implemented the bisection method in our calculation. In all the EoSs used here, atleast $20$ nodes lie well within $f < 50$ kHz. The computational findings will be discussed in detail in the following subsections. It is worth noting that the eigenfrequencies in section \ref{subsec:radosc} were angular frequencies ($\omega$), but we consider linear frequencies ($f$) where $\omega = 2\pi f$ in our results, since it is more prevalent in asteroseismology. Table. \ref{tab:20nodes} lists the frequencies of the 20 lowest radial oscillation nodes, measured at 1.4 M$_\odot$, for our selection of EoSs.

\begin{table*}
\label{tab:20nodes}
\centering
\begin{tabular}{|c|ccccc|ccc|}
\hline
\multirow{2}{*}{\begin{tabular}[c]{@{}c@{}}Order\\n\end{tabular}} & \multicolumn{5}{c|}{Hadronic EoSs}    & \multicolumn{3}{c|}{Quarkyonic EoSs}  \\ 
\cline{2-9}
                                                                  & DD2   & DDME2 & IOPB  & NL3   & SLy4  & Q1    & Q2    & Q3                   \\ 
\hline
0                                                                 & 3.11  & 3.25  & 2.87  & 2.59  & 3.01  & 5.61  & 4.95  & 3.45                 \\
1                                                                 & 6.43  & 6.52  & 6.15  & 5.56  & 6.96  & 8.03  & 8.11  & 7.45                 \\
2                                                                 & 8.65  & 8.56  & 8.28  & 7.28  & 9.60  & 9.63  & 10.21 & 9.18                 \\
3                                                                 & 9.36  & 9.31  & 8.99  & 8.30  & 10.81 & 11.75 & 11.97 & 10.71                \\
4                                                                 & 11.48 & 11.47 & 11.30 & 10.57 & 12.33 & 14.35 & 14.21 & 12.64                \\
5                                                                 & 13.18 & 13.13 & 12.74 & 11.17 & 14.59 & 15.88 & 15.95 & 14.38                \\
6                                                                 & 14.43 & 14.42 & 13.95 & 13.15 & 16.54 & 17.51 & 17.86 & 15.79                \\
7                                                                 & 16.18 & 16.14 & 15.78 & 14.35 & 17.89 & 19.98 & 20.51 & 18.03                \\
8                                                                 & 18.14 & 18.13 & 17.51 & 15.77 & 19.97 & 21.95 & 22.15 & 19.77                \\
9                                                                 & 19.28 & 19.22 & 18.72 & 16.97 & 22.08 & 24.39 & 23.99 & 21.46                \\
10                                                                & 21.12 & 21.15 & 20.64 & 18.83 & 23.61 & 25.88 & 26.47 & 23.59                \\
11                                                                & 23.17 & 23.14 & 22.17 & 19.93 & 25.4  & 27.71 & 28.49 & 25.50                \\
12                                                                & 24.40 & 24.31 & 23.62 & 21.48 & 27.59 & 30.25 & 30.23 & 27.19                \\
13                                                                & 26.03 & 26.06 & 25.22 & 23.38 & 29.41 & 32.53 & 32.52 & 29.03                \\
14                                                                & 28.15 & 28.24 & 27.27 & 24.38 & 30.97 & 34.18 & 34.66 & 30.79                \\
15                                                                & 29.65 & 29.53 & 28.30 & 26.02 & 32.99 & 35.99 & 36.69 & 32.67                \\
16                                                                & 31.07 & 31.09 & 30.38 & 27.46 & 35.15 & 38.34 & 38.67 & 34.53                \\
17                                                                & 33.08 & 33.13 & 31.89 & 28.93 & 36.72 & 40.75 & 40.84 & 36.36                \\
18                                                                & 34.69 & 34.54 & 33.44 & 30.29 & 38.55 & 42.44 & 42.93 & 38.32                \\
19                                                                & 36.15 & 36.24 & 34.98 & 31.92 & 40.63 & 44.34 & 44.88 & 40.21  \\             
\hline
\end{tabular}
\caption{Frequencies $f_n$ (in kHz) of the 20 lowest radial oscillation modes for our choice of 5 hadronic and 3 quarkyonic unified EoSs taken at 1.4 M$_\odot$.}
\end{table*}

\begin{figure}
    \centering
    \includegraphics[scale=0.55]{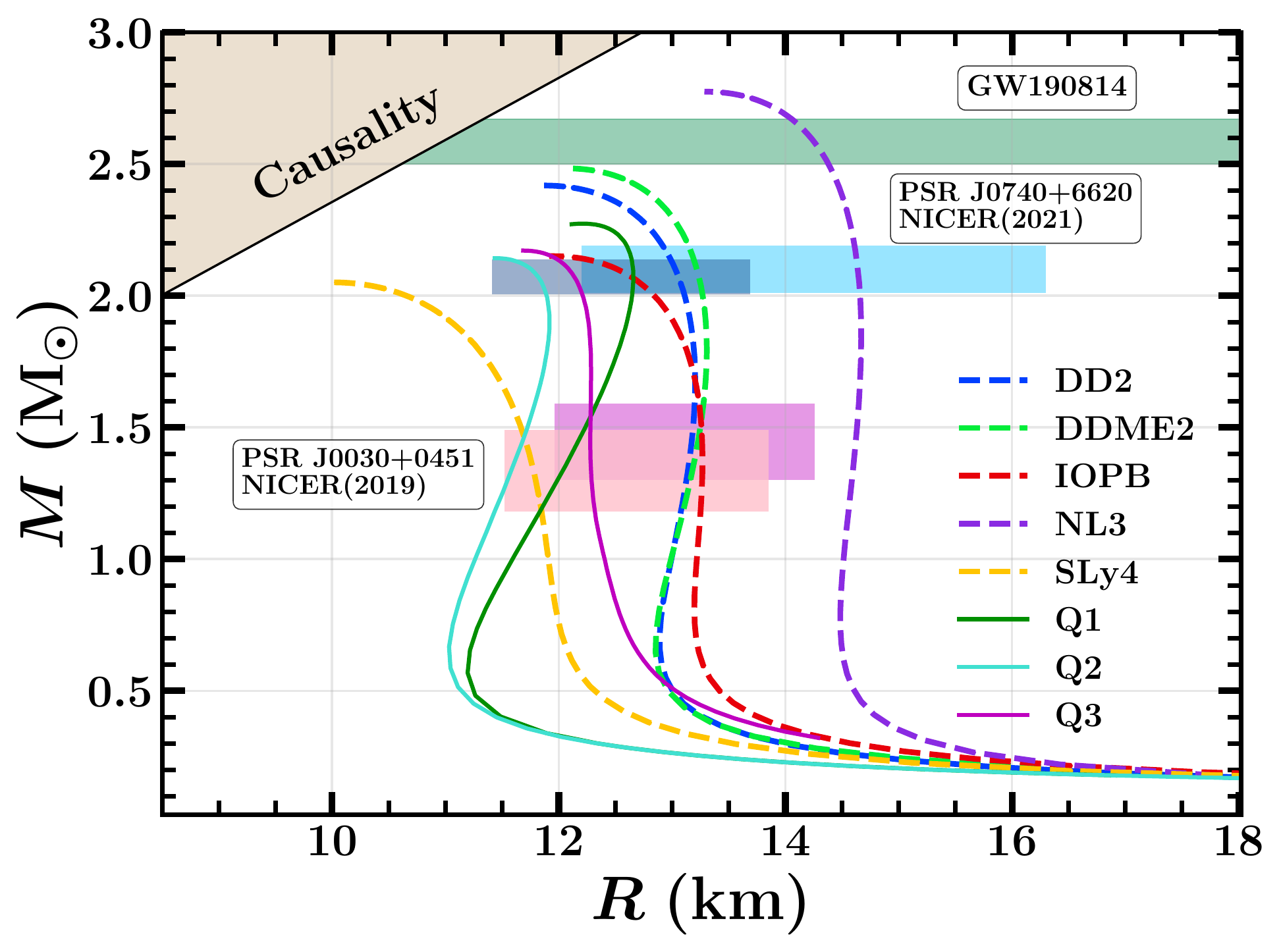}
    \caption{NS mass-radius relation for our choice of EoSs. The lower pink and magenta boxes show the constraints on mass and radius from the 2019 NICER data of PSR J0030+0451 by Riley and Miller \cite{Miller_2019,Riley_2019}. The upper dark and light blue boxes show the 2021 NICER data with X-ray Multi-Mirror observations of PSR J0740+6620 \cite{miller2021radius}. Observational mass data from merger event GW190814 is shown as a horizontal green band \cite{2020ApJ...896L..44A}. The upper left region is restricted so that the speed of sound in the object does not surpass the speed of light in vacuum (causality).}
    \label{fig:MR}
\end{figure}

\subsection{$M-R$ relation}
\label{subsec:mr}
For an EoS, computations begin at the star's centre with the initial parameter central density $\rho_c$ as input and continue solving the TOV equations \eqref{eqn6}) and \eqref{eqn5} until $P < 0$, to obtain the radius $R$ and mass $M$ of a NS. We repeat the procedure by changing $\rho_c$ in equal increments and performing the same calculations to obtain the Mass-Radius ($M-R$) plot. Figure \ref{fig:MR} shows the Mass - Radius plot for the eight EoSs considered in this work.

Neutron star Interior Composition Explorer (NICER) helps in the study of exotic matter and NS composition. In 2019 from the analysis of the NICER data of PSR J0030+0451 by Miller et al. \cite{Miller_2019} and Riley et al. \cite{Riley_2019}, gave the measurements of mass $M$ and radius $R$ as $M=1.44^{+0.15}_{-0.14}$ M$_\odot$; $R=13.02^{+1.24}_{-1.06}$ km and $M=1.34^{+0.15}_{-0.16}$ M$_\odot$; $R=12.71^{+1.14}_{-1.19}$ km respectively. These are shown as lower pink and magenta boxes in figure \ref{fig:MR}. 
Cromartie et al. \cite{cromartie2020relativistic} and Antoniadis et al. \cite{anto} through data from radio observations from PSR J0740 + 6620 calculated the NS mass as $M=2.14^{+0.1}_{-0.09}$ M$_\odot$ and $M=2.01^{+0.04}_{-0.04}$ M$_\odot$, respectively. While Fonseca et al. \cite{Fonseca_2021} calculated the mass of this pulsar using a model-averaged estimation. This gives the lower limit for the maximum mass of the NS as $M=2.08^{+0.07}_{-0.07}$ M$_\odot$. The  NICER data and X-ray Multi-Mirror (XMM) Newton data of the millisecond pulsar PSR J0740+6620 in 2021 by Miller et al. gave the radius to be $R=13.7^{+2.6}_{-1.5}$ km \cite{miller2021radius} while Riley et al. calculated it to be $R=12.39^{+1.30}_{-0.98}$ km \cite{Riley_2021}. These are represented by the upper dark and light blue horizontal boxes in the figure. The figure also has shown the data from the GW detection of a black hole and a compact object merger, GW190814 \cite{2020ApJ...896L..44A}, which tells us the secondary mass of the compact object to be $M=2.59^{+0.08}_{-0.09}$ M$_\odot$, depicted by the green horizontal band .

Among the quarkyonic EoSs, for a fixed $L$ and $n_{t}$ values one having higher $\Lambda$ has higher maximum mass and a more significant bending towards large radii, due to a higher peak in sound speed. A smaller transition density $n_{t}$ and a larger $L$ also leads to higher maximum mass and larger radii. As already verified by the authors in refs. \cite{PhysRevD.102.023021,PhysRevC.55.540,kumar2018new,typel2010composition,PhysRevC.DDME2,refId0}, these EoSs are consistent with the astrophysical data from pulsars and Gws and thereby are a good choice for our radial oscillation calculations. 

\begin{figure*}
    \centering
    \includegraphics[width=\linewidth]{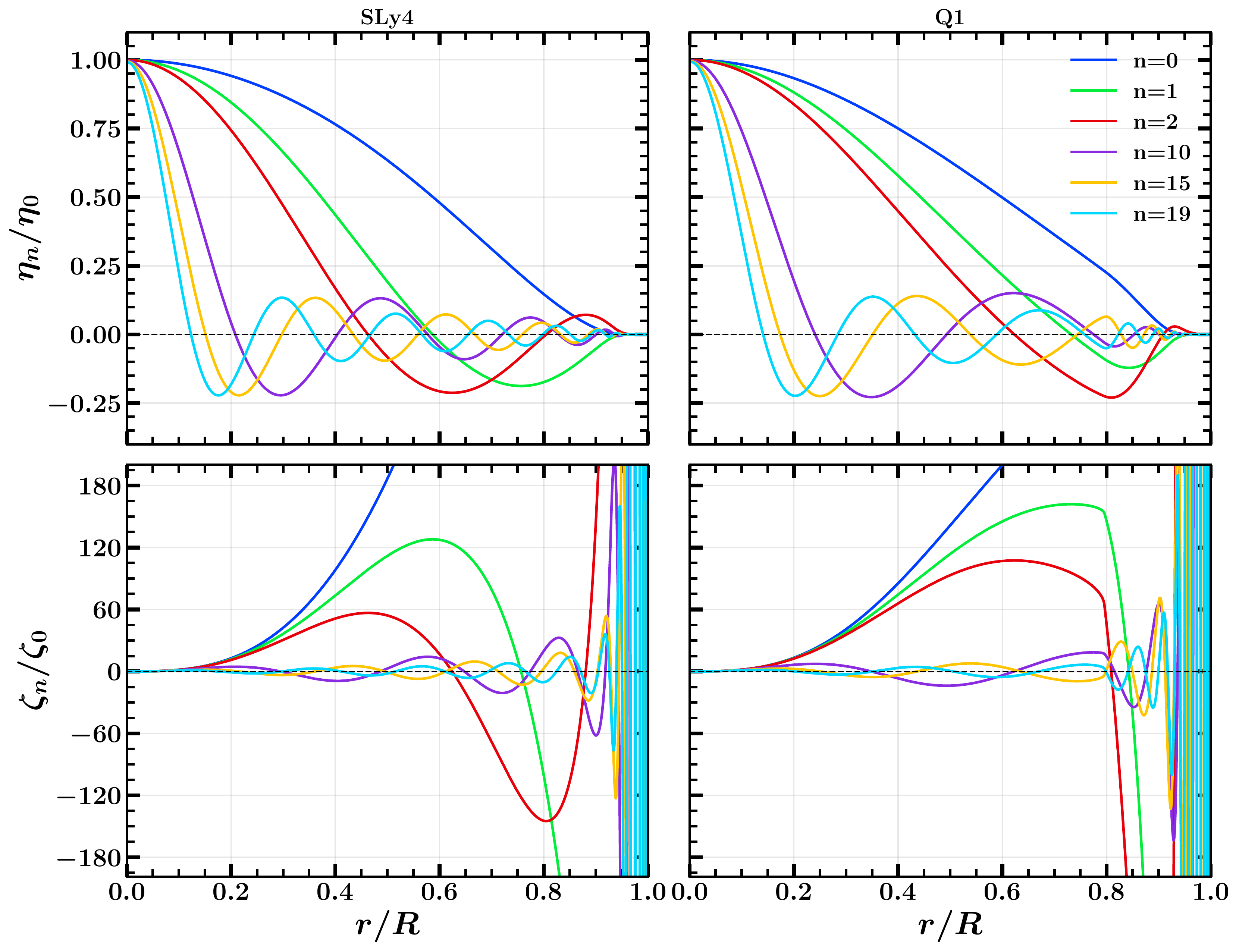}
    \caption{{\it Top panel:} $\eta/\eta_0$ as a function of dimensionless radial coordinate $r/R$ for low ($n=0,1,2$) and highly excited modes ($n=10,15,19$). {\it Bottom panel:} Same as before but for $\zeta/\zeta_0$. In both the panels, the left corresponds to a NS with SLy4 EoS and the right corresponds to that with Q1 EoS at 1.4 M$_\odot$. $\eta_0$ and $\zeta_0$ are the values of respective functions at the smallest radial coordinate $r_{\text{min}}$.}
    \label{fig:func-r}
\end{figure*}

\begin{figure*}
    \centering
    \includegraphics[width=\linewidth]{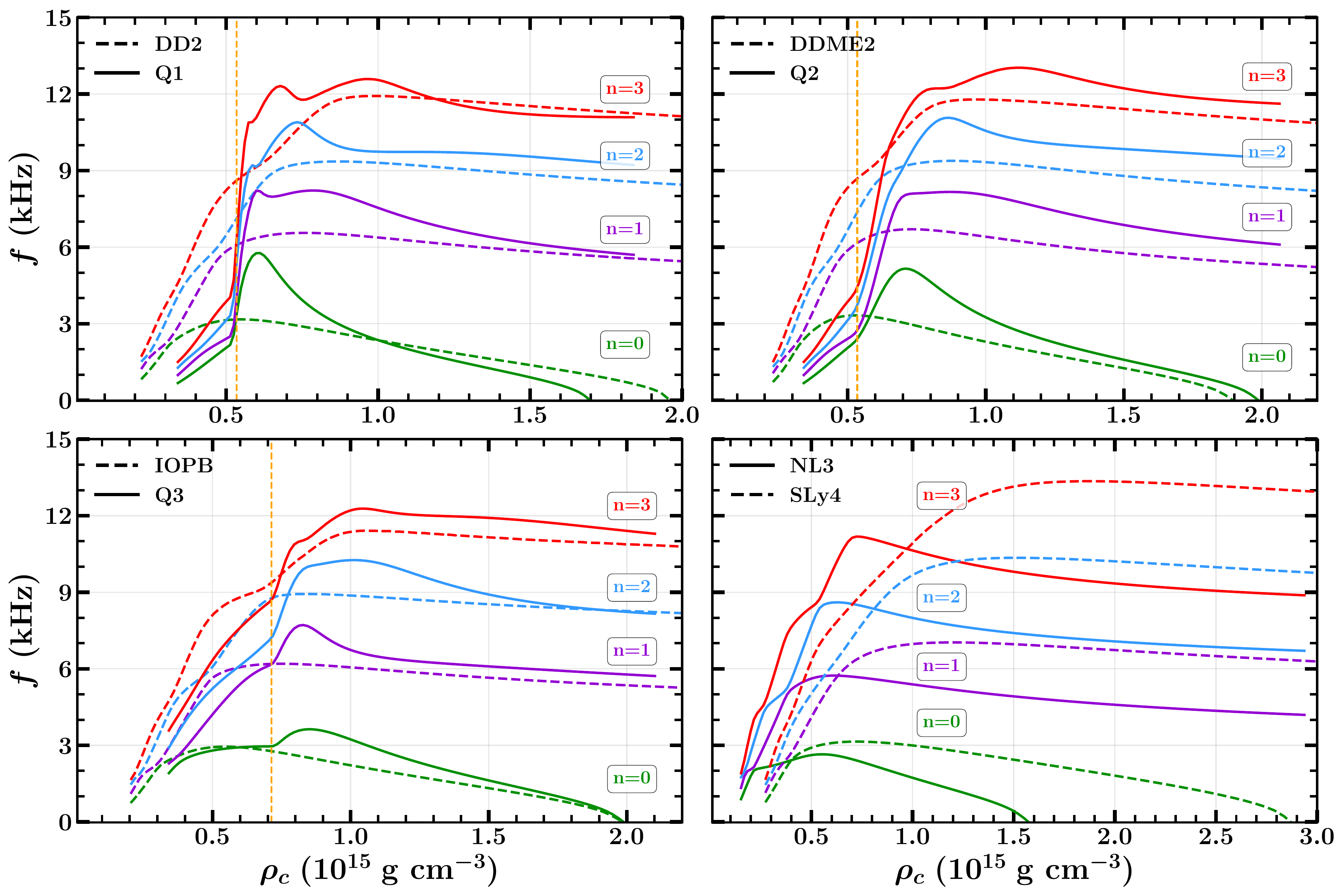}
    \caption{The first four radial modes as a function of central energy density. In the first three plots, we compare a hadronic EoS with a quarkyonic and in the last one, we compare the stiffest hadronic EoS NL3 with the softest SLy4. The vertical orange line in the first three plots marks the transition density ($n_t$) for the respective quarkyonic EoSs indicating the sudden change in frequencies.}
    \label{fig:f-rho}
\end{figure*}

\subsection{Radial oscillation modes}
\label{subsec:oscmode}
In figure \ref{fig:func-r}, we present the radial profile of $\eta_n$ and $\zeta_n$ for the f-mode ($n$=0) as well as excited p-modes ($n$=1,2,10,15,19) in 1.4 M$_\odot$ NS of SLy4 and Q1 EoS. Following the property of a Sturm-Liouville system, the eigenfunction $\zeta_n$ has exactly $n$ zeros (nodes) in the region $ 0<r<R$. While $\zeta$ is related to the radial displacement function by eq. \eqref{zeta}, and using eq.(\ref{rf},\ref{dp} \& \ref{DE1}) together, we get $\eta$ is associated to the Lagrangian variation of pressure by $\eta = -\Delta P e^{\lambda + 2\nu}$. Similar to $\zeta_n$, $\eta_n$ also has precisely $n$ zeros (nodes) between the centre and the stellar surface. We can observe the growing amplitude of $\zeta_n$ as the radial coordinate increases, whereas $\eta_n$ oscillates with a decaying amplitude and vanishes finally at the surface. Since $\eta_n$ is continuous, and so is $\Delta P$, it follows that the system always oscillates close to equilibrium \cite{PhysRevD.101.103003}. The radial functions $\zeta_n$ \& $\eta_n$ are smooth for hadronic SLy4 EoS, however close to $0.8R$ we observe abrupt changes for quarkyonic Q1 EoS. This abrupt behaviour is attributed to the jump in the adiabatic index $\gamma$ due to phase transition at $0.3$ fm$^{-3}$. For higher-order modes, some of the nodes move across the core-crust transition and lie in the crust $(0.9R \lesssim r \le R)$, where $\zeta_n$ changes signs rapidly with a large amplitude and appears as parallel vertical lines, see higher modes ($n$=10,15,19) in figure \ref{fig:func-r}. $\eta_n$, on the other hand, possesses a small amplitude in the crust.

In figure \ref{fig:f-rho}, we investigate the dependence of eigenfrequencies on central densities $\rho_c$ for our chosen EoSs, comparing a set of two EoSs at a time. We plot the first four radial modes and find that when density increases, regardless of EoS, we approach our stability limit. The instability point is defined by the presence of a zero eigenvalue for f-mode and corresponds to the central density $\rho_{\text{critical}}$ where the star approaches its maximum mass $M_{\text{max}}$ \cite{kokkotas2001radial}.

Moreover, with lower central density, an NS can be modelled as a homogeneous, non-relativistic body \cite{1983bhwd.book.....S,1992A&A...260..250V, 1977ApJS...33..415A} with angular frequency that follows $\omega^2 \propto\rho(4\gamma-3)$ \cite{2022arXiv220609407L}. This implies that the frequency is only affected by density because $\gamma$ is fairly constant at lower densities. Thus, regardless of EoS, frequency approaches to zero when the star's central density is sufficiently low, as illustrated in figure \ref{fig:f-rho}.

Furthermore, we could see a series of `avoided crossings' between separate modes where the frequencies of two consecutive modes from different families repel each other when they approach one another \cite{gondek1999avoided, kokkotas2001radial} . We have two independent families of radial oscillation modes, one at the high density core and the other at the low density envelope, partitioned at the neutron drip density. This neutron drip is linked with any realistic EoSs, hence the `avoided crossings' phenomenon, as shown in the figure, is likewise a feature of such EoSs. At the point of the `avoided crossings', the solution to the eigenvalue problem shifts from being a standing wave localized primarily in envelope to one that is localized mostly in the core \cite{gondek1999avoided}.

We also notice that quarkyonic EoSs have multiple peaks for higher order modes, see for example $n=3$ in figure \ref{fig:f-rho}. Because quarkyonic EoSs have a narrow peak of adiabatic index above the transition density. This local peak is a fine interior structure sensitive to higher order modes. This unique feature in higher order modes could be a strong indication of phase transition, if observed.

\begin{figure}
    \centering
    \includegraphics[scale=0.5]{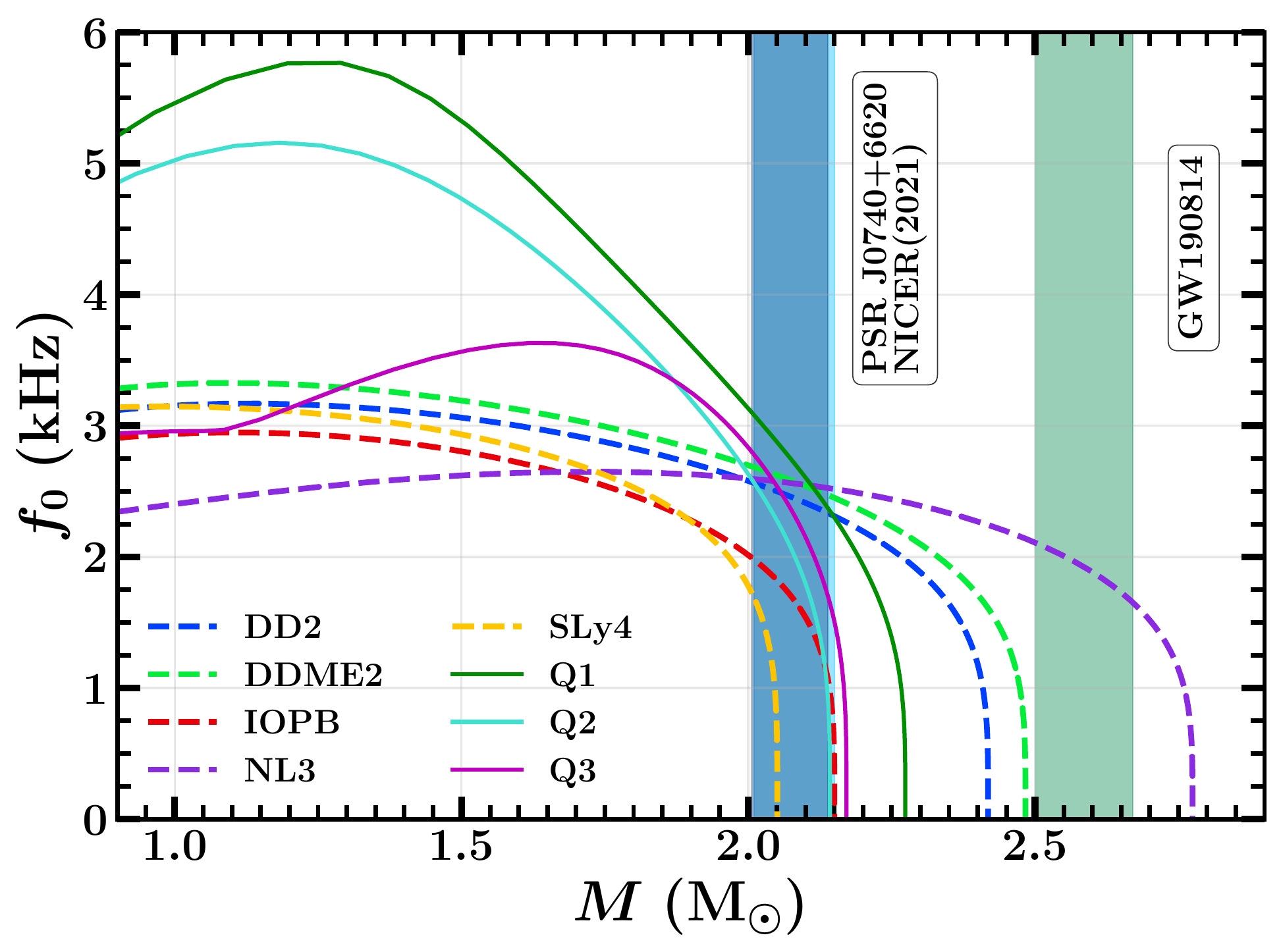}
    \caption{f-mode frequency vs mass are shown for our choice of EoSs. The blue vertical bands represents the 2021 NICER data with X-ray Multi-Mirror mass observations of PSR J0740+6620 \cite{miller2021radius} and the green vertical band represents the mass data from the GW190814 merger event \cite{2020ApJ...896L..44A}.}
    \label{fig:f-M}
\end{figure}

In figure \ref{fig:f-M}, we display the NS's f-mode frequency vs mass $M$ to investigate the relationship between radial oscillation and stability more thoroughly. The curves for the quarkyonic EoSs follow the same trend as that of hadronic EoSs, reaching exactly zero at $M_{\text{max}}$ (from the $M-R$ curve). However, in the case of quarkyonic EoSs, the f-mode frequency increases rapidly near 1.4 M$_\odot$ at which quarks start to drip out from nucleons, but in the case of hadronic EoSs, it remains practically constant until it falls at $M_{\text{max}}$ \cite{PhysRevD.103.103003}. Beyond this limit, $\omega^2$ becomes negative, therefore with imaginary frequency, the star can no longer recover from minor radial perturbations and ultimately collapses into a black hole. So, this result is consistent with the stability condition $\partial M/\partial\rho_c > 0$ \cite{1983bhwd.book.....S,1965gtgc.book.....H}.

\begin{figure}
    \centering
    \includegraphics[scale=0.55]{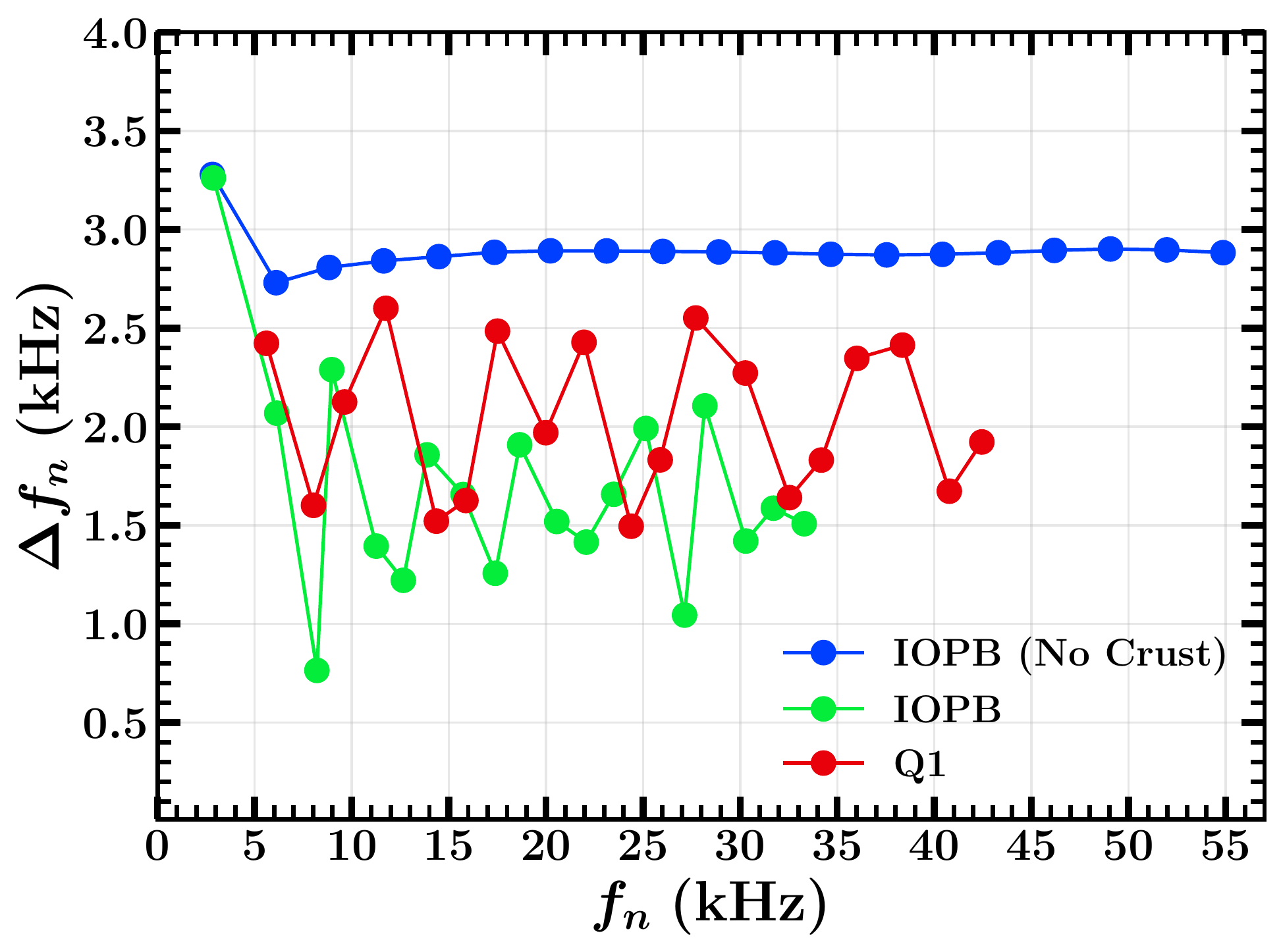}
    \caption{Large frequency separation vs frequency comparing a unified hadronic EoS IOPB to a quarkyonic Q1 at 1.4 M$_\odot$. Also, the nature of IOPB without crust (at the same mass) is shown to analyse how drastically the behaviour changes with the removal of the crust part from the EoS.}
    \label{fig:delf-f}
\end{figure}

Figure \ref{fig:delf-f} depicts the variation of the so-called large separation, the difference between consecutive nodes $\Delta f_n = f_{n+1}-f_n$, with $f_n$ calculated at 1.4 M$_\odot$, which aids in understanding the physics of the stars' interior and is a commonly used quantity in asteroseismology. When an EoS is considered without crust, the variation is smooth and consistent with Sagun et al. \cite{PhysRevD.101.063025}, with the first large separation $\Delta f_0$ being bigger than the rest. And as we decrease the central density from its critical point, where the star reaches its maximum mass, $\Delta f$ falls as well, see the difference between subsequent nodes in figure \ref{fig:f-rho}. But when we consider a unified EoS like IOPB and Q1, we obtain uneven fluctuations. This is because the inner crust,  the region in which the behaviour of adiabatic index is no longer monotonic, is believed to be characterized by complex structures collectively known as nuclear pasta. Outside nuclear pasta, adiabatic index is about $\gamma=4/3$ determined by relativistic electron gas, while inside the pasta, $\gamma \gtrsim2$. For the lowest order mode ($n=0$), the crust does not play a big role in radial oscillation, since it typically accounts for less than 10 per cent of the stellar radius, and the oscillation nodes lie deeply in the NS core, see figure \ref{fig:func-r}. Some of the oscillation nodes are in the crust for higher order modes, which is equivalent to lower order modes of NS without crust. In other words, in order to reach identical oscillation mode in a NS without crust, NS with crust need to have additional nodes in its crust. We also note that whenever a new node appears in the pasta region, $\Delta f$ shows a peak, meaning the $k$-th peak will have $k$ crustal nodes and ($n-k$) nodes inside the core, where $n$ is the total number of nodes. And the points between $k$-th and ($k+1$)-th peaks will also have $k$ and ($n-k$) nodes in the crust and the core, respectively. As a result, crust modulates the eigenfrequency significantly.

\subsection{Detectability}
\label{subsec:detect}
We may be able to learn about radial oscillation frequencies by examining the emission mechanism of a short gamma ray burst, as such oscillations influence SGRB from a hypermassive NS generated following a binary NS merger \cite{Chirenti_2019}. Radial oscillation can also couple with non-radial oscillation and amplify GWs  \cite{passamonti2006coupling, PhysRevD.75.084038}. Nonetheless, current GW detectors such as Advanced LIGO, Advanced Virgo, and KAGRA are projected to have a sensitivity of $\sim 2 \times 10^{−22} - 4\times10^{−24}$ strain/$\sqrt{\text{Hz}}$ at $\sim 20$ $\text{Hz} - 4$ $\text{kHz}$ \cite{PhysRevD.99.102004,abbott2020prospects}. Even third-generation ground-based detectors, such as the 40 km long Cosmic Explorer, may have a sensitivity under $10^{−25}$ strain/$\sqrt{\text{Hz}}$ above few kHz \cite{Abbott_2017}, while the underground 10 km long Einstein Telescope is expected to have a sensitivity $>3\times10^{−25}$ strain/$\sqrt{\text{Hz}}$ at 100 Hz and $\sim 6\times10^{−24}$ strain/$\sqrt{\text{Hz}}$ at $\sim 10$ kHz \cite{Punturo_2010}. As a result, neither the current nor the next generation of GW detectors could achieve the requisite sensitivities at the full frequency range in our study. However, detector sensitivities can be considerably increased by optical reconfiguration or with the use of advanced quantum techniques \cite{PhysRevD.99.102004,2019LRR....22....2D}. This could help us reach our target level, or we could have to wait for the fourth generation \cite{bora2021radial}.

\section{Conclusion}
\label{sec:con}
In this study, we investigated radial oscillations of NS while considering eight realistic EoSs based on the hadronic RMF, SHF and quarkyonic models \cite{Fortin,vishal,PhysRevD.102.023021}. Quarkyonic EoSs differ from hadronic EoSs by a peak in the speed of sound at which quark drip out of nucleons. We computed the mass and radius from the TOV equations for each EoS and verified that they are compatible with astrophysical observational evidence from pulsars and GWs \cite{cromartie2020relativistic,anto,2020ApJ...896L..44A,Miller_2019,Riley_2019,miller2021radius}. The radial oscillation equations were then solved considering infinitesimal adiabatic perturbation to calculate the modes of oscillation.

As we find highly excited modes, we show that $\eta_n(r)$ and $\zeta_n(r)$ changes sign exactly $n$ times within the star where we compare their behaviour in a hadronic star to a quarkyonic star to find that these functions change abruptly near $0.8R$ for the latter case due to the sudden jump in the adiabatic index following phase transition. We also investigate the behaviour of the first four radial modes as their central densities change. Our results for realistic EoSs resemble that of Kokkotas \& Ruoff \cite{kokkotas2001radial} and we also find that quarkyonic EoSs show multiple peaks at higher order. We then examine the nature of only the f-mode frequency with NS mass and observe that $f_0$ falls to zero at the maximum mass corresponding to its critical density. This outcome, regardless of EoS, is compatible with the static stability criterion $\partial M/\partial\rho_c > 0$. In $M-R$ relation, quarkyonic EoSs have a significant bend towards higher radius at higher mass. This unique feature also shows in f-mode frequency vs mass curves. Hadronic EoSs show a flat f-mode frequency however, f-mode frequency of quarkyonic EoS increases after the appearance of quark, forming a peak. And this peak frequency is higher than the f-mode frequency of any hadronic EoS. If we observe a f-mode frequency of $f>3.5$ kHz, it's a hint of strong crossover transition. This offers a chance to break the degeneracy of EoSs producing similar radius, e.g. SLy4 has identical radius but different f-mode frequency with Q1-3 for canonical mass NSs. In addition, we study how the nature of the large separation $\Delta f_n$ is smooth and evenly spaced for an EoS without crust. But its nature changes dramatically and becomes uneven with eigenfrequencies being squeezed within a shorter range when we include the crust. Finally, we discuss the possibility of detecting such radial frequencies when coupled with GWs, using our current and third-generation ground-based GW detectors.

Our comprehension of the internal structure of an NS will surely be further deepened by a reasonable application of this study in more realistic environments considering rotation, non-zero temperature and magnetic field \cite{panda2016radial}. Additionally, research on the potential for radial oscillation frequency detection and efforts to increase detector sensitivity will provide a chance for observational validation of our theory. We hope to explore these options in the near future.

\acknowledgments

The authors thank Grigoris Panotopoulos for his helpful comments and suggestions. BK acknowledges partial support from the Department of Science and Technology, Government of India with grant no. CRG/2021/000101. TZ is supported by the Department of Energy, Grant No. DE-FG02-93ER40756. SS is thankful to Yong Gao for his useful discussion about the numerical method for the calculation of eigenfrequencies.



\bibliographystyle{biblio}
\bibliography{main}



\end{document}